\documentclass[a4paper]{spie}  

\usepackage{amsmath,amsfonts,amssymb}
\usepackage{graphicx}
\usepackage[colorlinks=true, allcolors=blue]{hyperref}

\title{Quantum hydrodynamics for nanoplasmonics}

\author[]{Giovanni Manfredi}
\author[]{Paul-Antoine Hervieux}
\author[]{Fatema Tanjia}

\affil[]{Universit\'e de Strasbourg, CNRS, Institut de Physique et Chimie des Mat\'eriaux de Strasbourg, UMR 7504, F-67000 Strasbourg, France}

\authorinfo{Further author information: (Send correspondence to Giovanni Manfredi)\\Giovanni Manfredi: E-mail: giovanni.manfredi@ipcms.unistra.fr, Telephone: +33 (0)3 88 10 72 14}

\begin{document}
\maketitle

\begin{abstract}
Quantum effects play a significant role in nanometric plasmonic devices, such as small metal clusters and metallic nanoshells. For structures containing a large number of electrons, ab-initio methods such as the time-dependent density functional theory (TD-DFT) are often impractical because of severe computational constraints. Quantum hydrodynamics (QHD) offers a valuable alternative by representing the electron population as a continuous fluid medium evolving under the action of the self-consistent and external fields. Although relatively simple, QHD can incorporate quantum and nonlinear effects, nonlocal effects such as the electron spillout, as well as exchange and correlations.
Here, we show an application of the QHD methods to the plasmonic breathing oscillations in metallic nanoshells.  We illustrate the main advantages of this approach by comparing systematically the QHD results with those obtained with a  TD-DFT code.
\end{abstract}


\section{INTRODUCTION}
\label{sec:intro}

Small nanometric metallic or metal-like objects display a number of collective resonances, most notably the localized surface plasmon (LSP), which have attracted considerable interest over the last few decades. These resonances are the result of the collective motion of the electrons under the influence of an external electromagnetic field (laser). They can only be understood by taking into account the self-consistent internal fields created by the electrons themselves in response to the external excitation.

Theoretically, the electron response in metal nano-objects can be investigated at different levels of approximations. For small systems (say, less than a hundred electrons), ab-initio methods such as the time-dependent density functional theory (TD-DFT) are usually the preferred choice, but they become too costly, in terms of run time and memory storage, for larger nano-objects. Thus, for large systems, many recent investigations have relied on simpler methods, based on improvements of the classical Mie theory with quantum and nonlocal effects added ad-hoc\cite{Esteban2012}.

In between these two approaches, quantum hydrodynamics (QHD) \cite{Manfredi2001, Manfredi2005, Bonitz2018} describes the electron response using a small number of macroscopic fluid-like equations. QHD goes beyond the classical Mie theory by incorporating (at least to some lower order) such crucial features as nonlocal, nonlinear, quantum, and exchange-correlation effects. It is an appealing level of description for nanoplasmonics applications because it is light enough to allow the study of large nano-objects, but detailed enough to capture most of the key physical effects. Recent studies have shown the potential of QHD for quantum nanoplasmonics \cite{Ciraci2013,Ciraci2016}, but they are often restricted to the linear electron response and use commercial software packages \cite{Toscano2015}.
The strongly-excited nonlinear regime was investigated in several studies performed in our group \cite{Crouseilles2008,Haas2009,Hurst2014,Hurst2016}.

Most investigations of plasmonic modes have focussed on the LSP, which is a dipolar mode, because it is the easiest to excite in experiments where the laser wavelength is much larger than the size of the nano-object. Here we will present some preliminary results on a different family of modes, namely plasmonic ``breathing" oscillations, which are spherically symmetric monopolar modes. Although difficult to excite by optical means because of their symmetry \cite{Krug2014}, breathing modes can be driven and detected through electron energy loss spectroscopy (EELS) \cite{Schuler2016,Schmidt2012}.

\section{QHD MODEL}
\label{sec:model}

As a minimal QHD model to study plasmonic breathing modes, we consider a spherically-symmetric nano-object where all quantities depend only on the radial coordinate $r$ and the time $t$. The ion lattice is represented by a uniform continuous positive charge density (jellium), whereas the electrons are described by the following set of fluid equations (atomic units are used hereafter):
\begin{eqnarray}
&& \frac{\partial n}{\partial t} + {1 \over r^2}\frac{\partial}{\partial r}\left(n u r^2 \right)=0 \,, \label{eq:continuity}\\
&& \frac{\partial u}{\partial t}+u\frac{\partial u}{\partial r}  = \frac{\partial V_H}{\partial r}+\frac{1}{2}\frac{\partial}{\partial r} \left(\frac{\Delta_r\sqrt{n}}{\sqrt{n}}\right)-
{1 \over n}\frac{\partial P}{\partial r}
-\frac{\partial V_{X,C}}{\partial r}, \label{eq:euler}\
\end{eqnarray}
where $n$ is the electron number density, $u$ is the mean radial velocity, $P$ is the (isotropic) pressure, $V_{X,C}$ is the exchange and correlation potential, and $V_H$ is the Hartree potential obtained from Poisson's equation:
\begin{equation}
{1 \over r^2}\frac{\partial}{\partial r}\left( r^2 \frac{\partial V_H}{\partial r}\right) = 4\pi \left(n-n_i \right) \label{eq:poissonfluid},
\end{equation}
where $n_i$ is the density of the spatially-uniform ion jellium.
The second term on the right-hand side of Eq. \eqref{eq:euler} is the so-called Bohm potential, which contains quantum effects to lowest order, and can be shown to correspond to the first gradient correction to the electron kinetic energy.

For the exchange potential, we use the standard local density approximation (LDA):
\begin{equation}\label{eq:Vx}
V_X [n]= -\frac{(3\pi^2)^{1/3}}{\pi}\, n^{1/3},
\end{equation}
and for the correlations we employ the functional proposed by
Brey et al.\cite{Brey1990}, which yields the following correlation potential:
\begin{equation}\label{eq:Vc}
V_{C}[n] = -\gamma \ln \left(1+\delta n ^{1/3}\right),
\end{equation}
with $\gamma = 0.03349$ and $\delta = 18.376$.
We further assume the electron temperature to be much lower than the Fermi temperature of the metal, so that the pressure can be approximated by that of a fully degenerate electron  gas:
\begin{equation}\label{eq:pressure}
P= {1 \over 5}(3\pi^2)^{2/3} n^{5/3}.
\end{equation}

With the aim of modelling  metallic nanoshells, the ion jellium density $n_i$ is chosen to be constant and equal to $n_0$ inside a spherical shell of internal radius $R_i$ and external radius $R_e$, and zero outside. We further define the nanoshell mean radius $R=(R_i+R_e)/2$ and the thickness $\Delta=R_e-R_i$. Assuming global charge neutrality, the total number of electrons inside the shell is:
\begin{equation}\label{eq:Nelectrons}
N = n_0\, \mathcal{V} =\frac{R_e^3-R_i^3}{r_s^3},
\end{equation}
where $\mathcal{V}$ is the volume of the nanoshell and $r_s$ is the Wigner-Seitz radius of the metal. Finally, the plasmon frequency can be written as: $\omega_p=\sqrt{3/r_s^3}$.
In the forthcoming sections, we will consider sodium clusters with $r_s=4$.

\section{NUMERICAL RESULTS -- GROUND STATE}

\label{sec:groundstate}
   \begin{figure} [ht]
   \begin{center}
   \includegraphics[height=6cm]{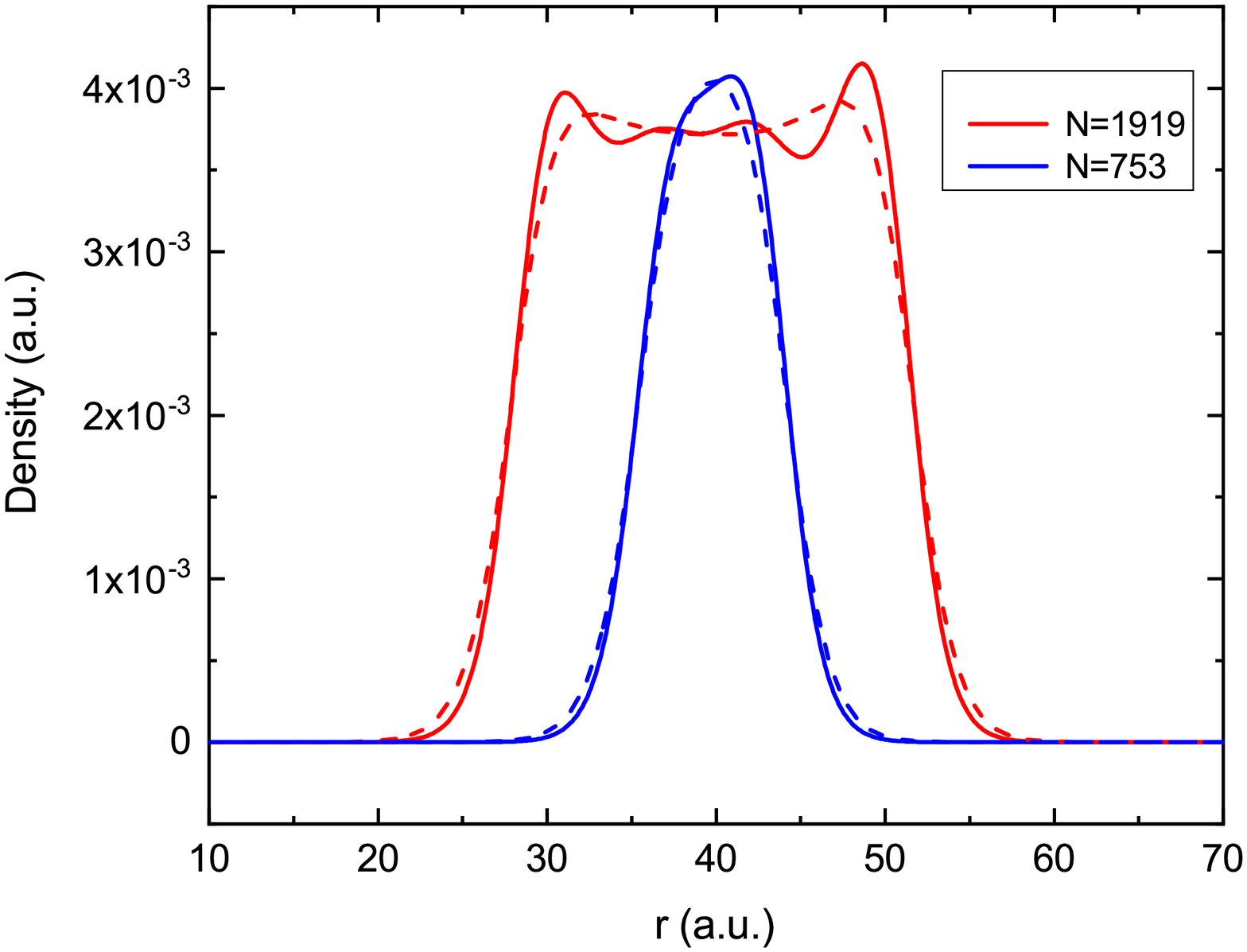}
   \includegraphics[height=6cm]{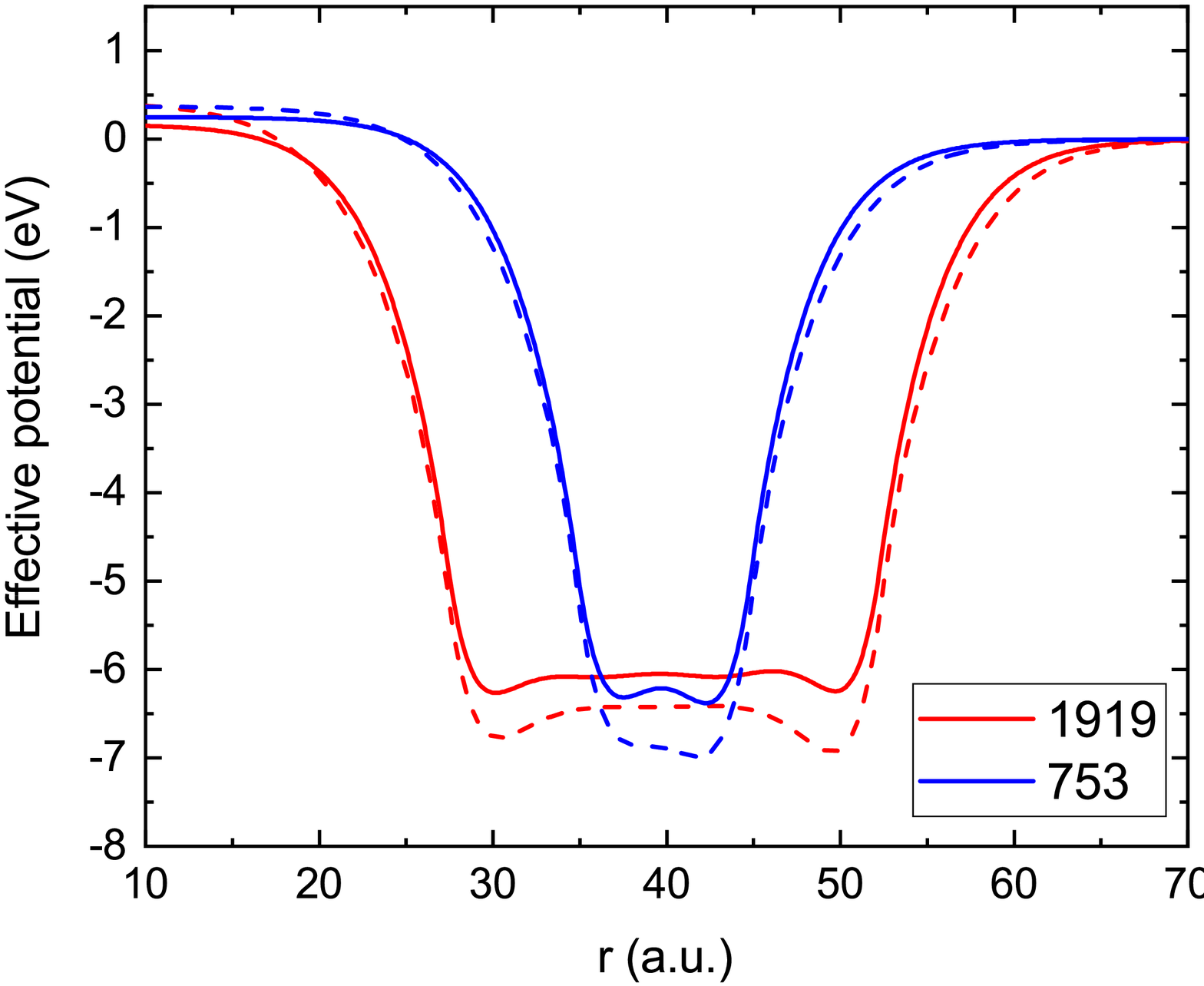}
   \end{center}
   \caption
   { \label{fig:gshydro}
Ground state of two typical Na nanoshells  containing $N=1919$ and $N=753$ electrons. Left panel: Electron densities computed using a DFT code (solid lines) and the QHD  approach (dashed lines). Right panel: Effective potentials for the same cases.
}
   \end{figure}

\label{sec:groundstate}
   \begin{figure} [ht]
   \begin{center}
   \includegraphics[height=6cm]{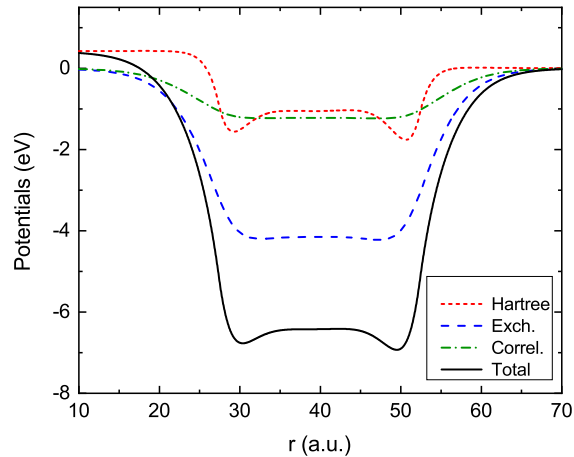}
   \end{center}
   \caption
   { \label{fig:gspoten}
Various potentials for the $N=1919$ case obtained with the QHD model: Effective total potential (solid line), Hartree potential (dotted line), exchange (dashed line) and correlation (dash-dot line) potentials.
}
   \end{figure}

Before considering the dynamical response of the system, we need to compute its ground state. This can be obtained as a stationary solution of Eqs. \eqref{eq:continuity}-\eqref{eq:euler} where we take $\partial/\partial t=0$ and $u=0$, and is calculated numerically using an iterative relaxation procedure \cite{Crouseilles2008}.

The electron density and effective potential profiles for two typical Na clusters containing respectively $N=1919$ and $N=753$ electrons are presented in Fig. \ref{fig:gshydro}. Both clusters have a central radius $R = 40$, but different thicknesses ($\Delta=25$ and $\Delta=10$).
On the same figure, we also show the density profiles computed using a standard DFT code. The agreement is very good, and even more so considering that the QHD results require no more than a few minutes runtime on a standard desktop computer. In particular, the nonlocal spillout effect is well described by the QHD method, even for the smaller structure ($N=753$), where the spillout is very prominent and the electron density is nowhere flat.
For the larger nanoshell ($N=1919$), the two overdensity ``ears" appearing at the internal and external radii in the QHD density profile are a lower-order quantum effect, a remnant of the well-known Friedel oscillations visible in the corresponding DFT profile.
It appears that (right panel of Fig. \ref{fig:gshydro}) the QHD method slightly overestimates the depth of the effective potential well.

Figure \ref{fig:gspoten} shows the various contributions to the effective potential for the larger nanoshell. The exchange potential constitutes the dominant contribution, with the Hartree and correlations parts being less than half as small.

\section{NUMERICAL RESULTS -- DYNAMICS}
\label{sec:linearresponse}

Having computed the ground state of the electron system, we need to perturb it slightly in order to induce some dynamical behavior. As we are interested in plasmonic breathing modes, the perturbation will also be spherically symmetric.
For the excitation, we use an instantaneous Coulomb potential applied at the initial time:
\begin{equation}
V_{ext}(r,t) = \frac{z}{r}\, \tau\,\delta(t),
\label{eq:excitation}
\end{equation}
where $\delta$ is the Dirac delta function, $z$ is a fictitious charge quantifying the magnitude of the perturbation, and $\tau$ is the duration of the pulse.

\begin{figure} [ht]
   \begin{center}
   \includegraphics[height=6cm]{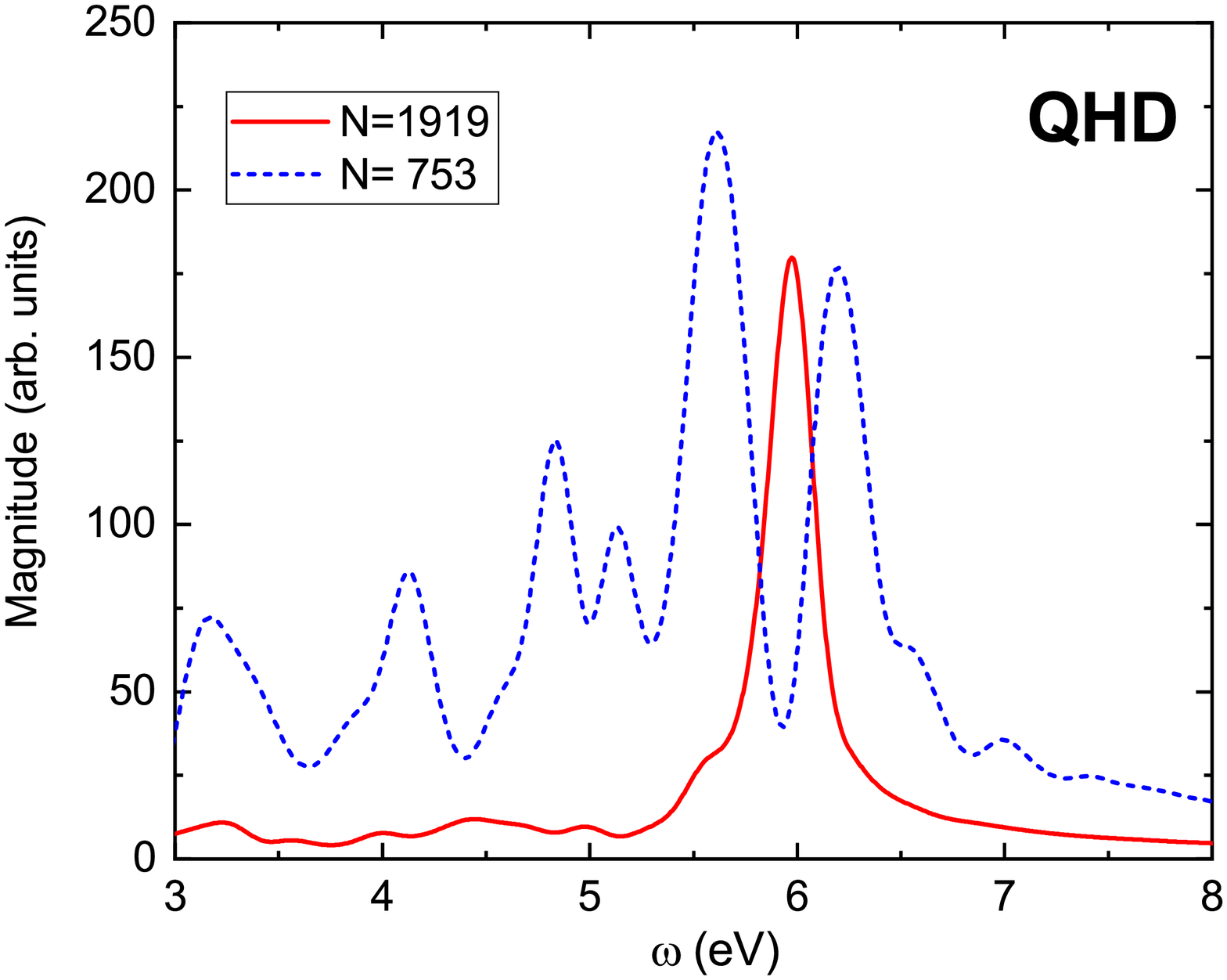}
 \includegraphics[height=6.cm]{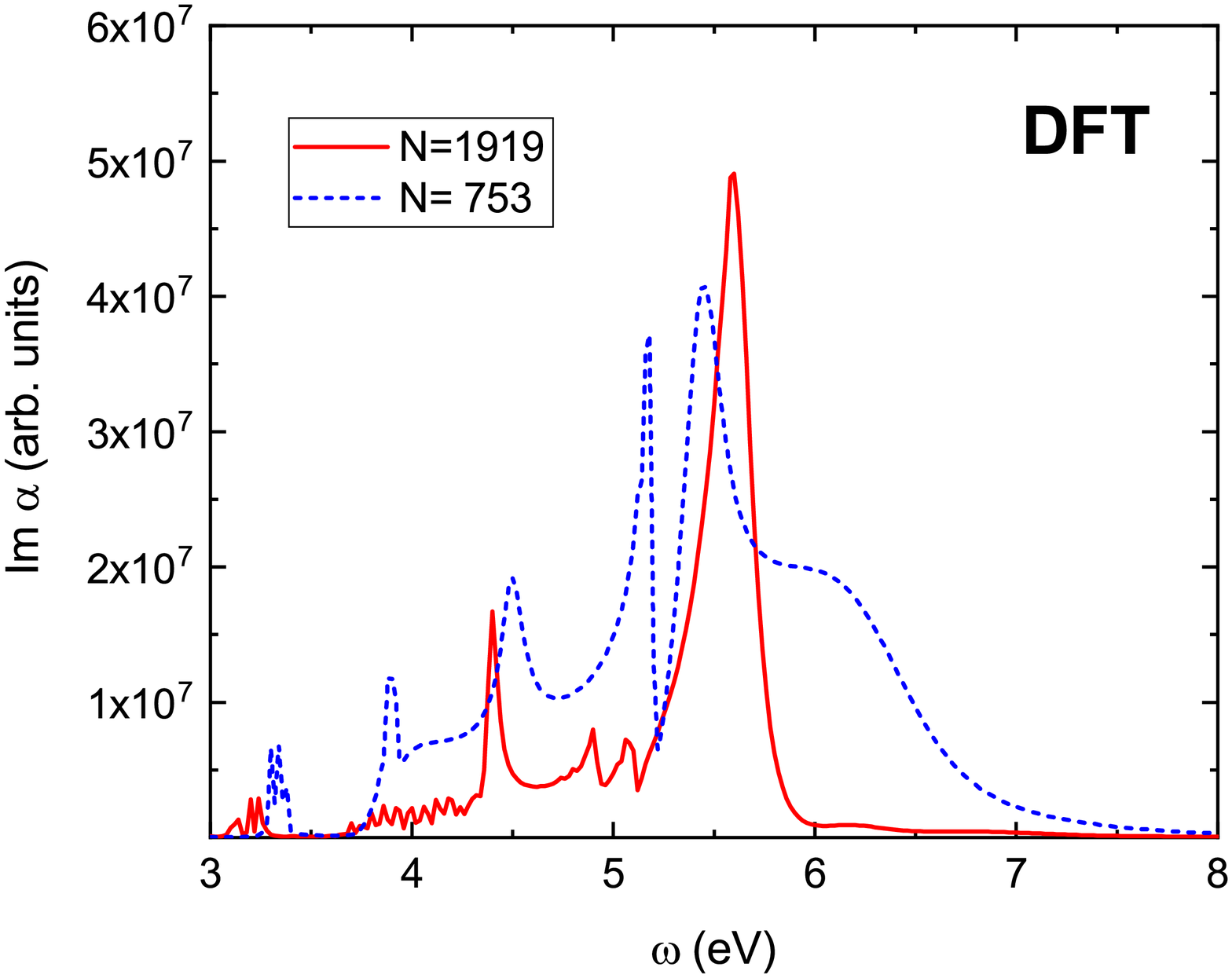}
   \end{center}
   \caption{ \label{fig:spectra}
Linear-response frequency spectrum  for two typical Na nanoshells containing $N=1919$ (red solid lines) and $N=753$ (blue dotted lines) electrons. Left frame: quantum hydrodynamics results; Right frame: TD-DFT results.
}
   \end{figure}

In order to analyze the linear response of the system, we study the evolution of the mean radius of the electron cloud, defined as:
\begin{equation}
\langle r \rangle (t)= {1\over N}\, \int_0^{\infty} r\, n(r,t) \,4\pi r^2 dr.
\label{eq:raverage}
\end{equation}
The Fourier transform of $\langle r \rangle$ in the frequency domain is shown in Fig. \ref{fig:spectra}, left frame. For the larger nanoshell ($N=1919$), the frequency spectrum shows a sharp peak near the plasmon frequency $\omega_p =\sqrt{3/r_s^3}= 5.89$~eV. This behavior is compatible with the computed ground-state density profile (Fig. \ref{fig:gshydro}), which displays a region of almost constant density in between the inner and outer radii.
In contrast, the spectrum of the smaller nanoshell ($N=753$) is much more fragmented and actually displays {\em two} principal peaks around the plasmon frequency, at about 5.6~eV and 6.2~eV, plus a number of smaller peaks at lower energies.

Still on Fig. \ref{fig:spectra} (right panel), we show the monopolar polarizability $\alpha$ computed with a linear-response TD-DFT code \cite{Maurat2009}, using the same parameters and exchange-correlation functionals as for the QHD simulations. The results compare rather favorably with the QHD ones. For $N=1919$, one dominant peak is observed at 5.6~eV, i.e. slightly redshifted compared to the pure plasmon frequency. This redshift can be understood in terms of dissipative phenomena (such as Landau damping, i.e. the coupling of the plasmon mode to single-particle modes) that are not included in the QHD description.
More interestingly, the spectrum of the smaller nanoshell ($N=753$) reveals the same two-peak structure also observed in the QHD simulations, now with frequencies $\approx 5.16 \,\rm eV$ and 5.45~eV, again slightly redshifted compared to corresponding peaks in the QHD spectrum.

The more complex spectrum observed for $N=753$ is probably due to the shape of the ground state electron density, which looks more like a bell curve  with no flat region inside the lattice jellium (see Fig. \ref{fig:gshydro}). In the case of such smooth density profile, the standard Mie theory of the localized surface plasmon is not applicable, as surface and volume plasmons are no longer well distinguished.

\label{sec:groundstate}

\section{CONCLUSIONS}
\label{sec:conclusion}

In this work, we have illustrated the main features of quantum hydrodynamics and described how this approach can be used to address some typical problems arising in nanoplasmonics. The main advantage of QHD is that it can be used to treat large systems (hundreds or thousands of electrons) with a relatively modest computational cost. Importantly, the most relevant effects for nanoplasmonics applications -- such as nonlinear and nonlocal effects -- can be included, at least to some degree of approximation. Its main limitation is that, unlike DFT, quantum effects arising from the discrete nature of the electronic energy levels cannot be taken into account, although some quantum features are retained to lowest order through the Bohm potential.

As a typical example, the QHD approach was used here to investigate the ground state and linear response of metallic nanoshells. Detailed comparison with a fully quantum TD-DFT code revealed a satisfactory accordance between the two approaches. These results are encouraging for future, more realistic studies of large nanoplasmonic structures using QHD.

\acknowledgments 
This project has received funding from the European Union's Horizon 2020 research and innovation programme under grant agreement No 701599 -- QHYDRO -- H2020-MSCA-IF-2015.

\bibliography{bibfile_qhd} 
\bibliographystyle{spiebib} 

\end{document}